\documentclass{aa}

\input psfig2.sty
\begin{document}
\title{A 1.4 GHz radio continuum and polarization survey\\
at medium Galactic latitudes:\\
I. Observation and reduction technique}

\author{B. Uyan{\i}ker, E. F\"urst, W. Reich, P. Reich, and R. Wielebinski}

\institute{Max--Planck--Institut f\"ur Radioastronomie,
        Postfach 2024, 53010 Bonn, Germany}

\thesaurus{22 (02.16.2; 03.13.2; 03.20.5; 04.19.1; 10.19.3; 13.18.3) }

\offprints{W. Reich}

\date{Received 26 December 1997 / Accepted 27 April 1998}

\titlerunning{A 1.4 GHz radio continuum and
polarization survey at medium Galactic latitudes}
\authorrunning{B. Uyan{\i}ker et al.}

\maketitle

\begin{abstract}
A radio continuum survey at medium Galactic latitudes with the Effelsberg
100-m telescope is being carried out at a centre frequency of 1.4~GHz in total
power and linear polarization. Areas up to $\pm 20\degr$ of Galactic latitude
are now being observed at a sensitivity of 15~mK T$_\mathrm{B}$ in
total intensity and 8~mK T$_\mathrm{B}$ in linear polarization with an
angular resolution of $9\farcm 35$. This paper describes the observing
and reduction technique applied which results in absolutely calibrated
maps. The methods are illustrated by examples of images from the survey.

\keywords{Polarization -- Methods: data analysis -- Techniques: polarimetric --
Surveys -- Galaxy: structure -- Radio continuum: ISM}
\end{abstract}

\section{Introduction}
The interaction between the Galactic  magnetic field and interstellar clouds
leads to a variety of radio emitting structures, shells,  filaments, and
loops. Many of these features are linearly polarized. They extend up to high
Galactic latitudes but they are very faint, an order of magnitude fainter than
the large-scale diffuse Galactic emission.  However, the study of these
structures may  provide important clues for the  understanding  of the
``disk--halo connection''.  The interpretation favours chimneys, fountains,
Parker loops or superbubbles created by OB associations to transfer material
from the disk into the halo.  They will deform and compress the Galactic
magnetic field. If these structures account for the rather smooth nonthermal
background, they have to be sufficiently numerous and large.

The Galactic region above $b = +4\degr$ and below $b = -4\degr$, where the
just mentioned structures play an important role, has never been studied in a
systematic way. Large-scale surveys covering this area either suffer from low
angular resolution or they miss the diffuse emission as it is the case
for the 1.4~GHz VLA-survey (Condon et al.\  \cite{condon96}). A combination of
1.4~GHz data from the Effelsberg telescope with high angular resolution
data from the VLA is of particular importance to separate compact
background sources from faint extended Galactic structures. An
example has already been shown and discussed by F\"urst et al.
(\cite{fuerst+98}).

One of the milestones in the research of Galactic magnetic fields is due to
Brouw \& Spoelstra (\cite{brouw76}) and Spoelstra (\cite{spoelstra84}).
They have surveyed the northern sky in linear polarization with the
Dwingeloo 25-m telescope at 1.411~GHz at an angular resolution of about
half a degree. They also compiled maps of the linear polarization at
frequencies between 408~MHz and 1.411~GHz. At 2.7~GHz Junkes et al.
(\cite{junkes+87}) published a survey of the linear polarization of
the Galactic plane ($ \vert b \vert \leq 1\fdg 5$ and $ 4\fdg 9 \leq
\ell \leq 76\degr$). Recently, the southern Galactic plane ($ \vert b
\vert \leq 5\degr$) was surveyed at 2.4~GHz by Duncan et al.
(\cite{duncan95}). Aside from the Dwingeloo data (see the discussion in
Sect.~5), all these polarization surveys are constrained to the
Galactic plane. On the low frequency side, at 327~MHz, Wieringa et al.
(\cite{wieringa+93}) observed small-scale variations in polarization
at high latitudes with the Westerbork synthesis telescope.

The knowledge of the magnetic field in the Galactic halo even in the
immediate vicinity of the Galactic plane is limited, because of the
lack of observations of the linear polarization at the proper angular
resolution and sensitivity. To fill this gap new observations at
1.4~GHz are being carried out to cover the entire Galactic plane at
medium Galactic latitudes ($\vert b \vert \leq 20\degr$) visible at
Effelsberg. This survey will reach a sensitivity at total intensity
close to the confusion limit ($\sim 15$~mk T$_\mathrm{B}$)
and an even higher sensitivity ($\sim 8$~mK T$_\mathrm{B}$)
at linear polarization.

We discuss the general survey parameters and observing method in
Sect.~2. An absolute calibration method for the total intensity data is
described in Sect.~3. The method developed to overcome the usual
problem of polarimetric observations, the instrumental polarization, is
demonstrated in Sect.~4. In Sect.~5 we propose a procedure
to adjust the polarization data to an absolute temperature scale.

\section{Observing method and data reduction}
The observations of the survey are carried out with the two-channel
1.3--1.7~GHz receiver installed in the primary focus of the Effelsberg
100-m telescope. The receiver is equipped with cooled HEMT amplifiers
recording the left- and right-hand circularly polarized component
simultaneously (see Schmidt \& Zinz\  \cite{schmidtzinz94}). These
amplifiers are of very high stability, and total intensity observations
are limited by confusion which is reached after about 2~s of
integration time. An IF-polarimeter converts the circularly polarized
component into linear components (Stokes parameter $U$ and $Q$). Stokes
parameter $I$ is obtained by simply adding both circularly polarized
components.

The method of observation is to scan each field along constant Galactic 
latitude as well as along constant Galactic longitude. Each field was 
measured at least twice. The telescope parameters relevant to the 
1.4~GHz survey are listed in Table~\ref{surpar}.

\begin{table}[htb]
\caption{Parameters for the survey}
\label{surpar}
\begin{tabular}{lr} \hline
\noalign{\smallskip}
Parameter & Value \\
\noalign{\smallskip}
 \hline\noalign{\smallskip}
Centre Frequency            & 1.4 GHz$^*$ \\
Bandwidth                   & 20 MHz$^*$ \\
Integration time per point       & 2 s \\
System temperature           & 26 K \\
Typical rms noise in total power & 15 mK \\
Typical rms noise in polarization & 8 mK \\
T$_\mathrm{B}$ / S               & $2.12\pm 0.02$ K/Jy \\
Telescope beamwidth          & $9\farcm 35 \pm 0\farcm 04$ \\
Typical scan length             & 10$\degr$ \\
Scan interval                  & 4$\arcmin$ \\
Scanning velocity           &4\arcmin /s\\
\noalign{\smallskip}
\hline
\end{tabular}\\[1ex]
$^*$ sometimes interference makes a shift of the centre frequency\\
and/or a reduction of the bandwidth necessary.
\end{table}

Daytime observations cause inacceptable distortions in both the total
intensity and the polarization data due to contribution of solar
emission to the far sidelobes of the telescope (Kalberla et al.\
\cite{kalberla+80}). Therefore all observations are carried out
at nighttime, when additional ionospheric Faraday rotation effects
are minimal and included by corrections inferred by observations of
calibration sources. Calibration sources were 3C286, 3C138 and 3C48, 
where 3C286 served as the primary calibrator both in total intensity and
polarization. The scale accuracy for total and polarized intensities is
better than 5\%.  As seen from their coordinates, calibration sources have 
large angular separations on the sky, so that at least one of them is
accessible at all times. Their polarization properties are listed in
Table~\ref{cali} and are taken from the list of Tabara \& Inoue
(\cite{tabarainoue80}).

\begin{table}[htb]
\caption{Major calibration sources}
\label{cali}
\begin{tabular}{lrrccl}
\hline\noalign{\smallskip}
Source &$\ell$~~ &  $b$~~& $S_\mathrm{21cm}$/Jy & \% Pol. & PA \\
\noalign{\smallskip}
\hline\noalign{\smallskip}
3C286  &  56.5  & +80.7 & 14.4 &  9.3 & 32 \\
3C138  & 187.4 &$-$11.3 & ~~9.5 &  7.9 & 169.9 \\
3C~ 48 & 134.0 &$-$28.7 & 15.9 &  0.4 & 139.7 \\
\noalign{\smallskip}
\hline
\end{tabular}
\end{table}

\begin{figure}[htb]
\psfig{file=1502f1.ps,width=8.8cm,bbllx=60pt,bblly=180pt,bburx=522pt,bbury=570pt,clip=}
\caption[ ]{1.4 GHz beam map covering an area of 1\fdg1 $\times$ 1\fdg1.
Contours run in steps of 3 dB. The lowest contour is at $-$30~dB and the highest
contour is at $-$3~dB of the maximum.}
\label{beam}
\end{figure}

In Fig.~\ref{beam} we show a beam map at 1.4~GHz observed in the
Az/El-system on 3C123, which at that frequency shows no measurable
variation with elevation. Compact background sources in the field of 3C123
have been removed. The dynamic range exceeds 30~dB. The maximum
of the first sidelobes, which are enhanced by the four subreflector support
legs, are at a level of about $-$20~dB. The cross-polarization lobes of Stokes
parameters $U$ and $Q$ are dominated by a main beam component, which 
varies with parallactic angle. A correction of that instrumental response is 
described below.

The first stage of the data reduction made use of the standard
``Toolbox'' procedure (CONT2, von Kap-herr\  \cite{kapherr77}) for continuum 
and polarization observations with the Effelsberg telescope. From the tabulated 
scans calibrated maps were computed for all the four data channels (two times 
$I$, $U$ and  $Q$) recorded. A linear baseline using data points at the end of each 
scan was subtracted from each channel. All later stages of the data reduction are 
based on the NOD2 program package (Haslam\ \cite{haslam74}). Spiky data were
removed. Scanning effects were suppressed by using the method of
unsharp masking developed by Sofue \& Reich (\cite{sofuereich79}).
Finally, the two maps observed in orthogonal directions for each of the
Stokes parameters have been added using the PLAIT program (Emerson \&
Gr\"{a}ve\  \cite{emerson88}), which reduces ``scanning effects'' from 
individual distortions or by the baseline setting procedure in the case of an emission
structure located at the edge of a map by appropriate weighted addition of the Fourier 
transforms of both maps. The polarization $U$ and $Q$ maps were corrected for 
instrumental polarization as described in Sect.~4 before combining the maps observed
in orthogonal directions.

\section{Absolute calibration of the Effelsberg total intensity maps}
From the observation and the method of data reduction it is obvious
that neither the total intensity nor the polarization maps are on an
absolute temperature scale. The total intensity maps are calibrated to
an absolute scale using the 1.4~GHz northern sky survey by Reich
(\cite{reich82}) and Reich \& Reich (\cite{reichreich86}) carried
out with the Stockert 25-m telescope. This procedure has already been
described by Reich et al. (\cite{reich+90}) when calibrating the
Effelsberg 1.4~GHz Galactic plane survey. The Stockert survey has a scale
accuracy of 5\% and an uncertainty of the absolute zero-level of 0.5~K.

Briefly, the Effelsberg data and the Stockert data are convolved to an
angular resolution slightly exceeding that of the Stockert 1.4~GHz
survey (HPBW 36\arcmin) and the difference between the two data sets is
added to the Effelsberg data. This procedure also improves the original
Effelsberg maps since large-scale distortions by atmospheric
and ground radiation variations are removed.

However, still existing faint baseline effects in the low resolution
Stockert data are also added to the higher resolution Effelsberg data.
Such distortions were found to exist on small scales but not on the
large-scale background component, which is the missing component of the
Effelsberg maps. A modified version to calibrate the Effelsberg
maps absolutely was introduced:

\begin{itemize}
\item[$\bullet$] Convolve the Effelsberg map to the resolution of the
Stockert map (36\arcmin ). Decompose this map into a large-scale
component (Eff.back) and a map with small-scale structures using the
``background filtering method'' (Sofue \& Reich\  \cite{sofuereich79})
with a $3\degr$ Gaussian beam for smoothing.

\item[$\bullet$] Subtract the 2.8~K isotropic background component from the
Stockert data and convert the temperature scale from full beam into
main beam ($T_\mathrm{MB}/T_\mathrm{FB} = 1.55$).

\item[$\bullet$] Decompose  the Stockert data to obtain the corresponding
large-scale component (Sto.back) by using the same filtering
parameters as for the Effelsberg data.

\item[$\bullet$] Adjust Eff.back to Sto.back as described above,
namely, add the difference of both large-scale component maps to the
Effelsberg large-scale data and add the 2.8~K isotropic background
component. The resulting map is ES.back.

\item[$\bullet$] The final Effelsberg map (Eff.final) is given by:

\begin{center} Eff.final = Eff.orig + ES.back - Eff.back. \end{center}
\end{itemize}

\begin{figure}[h]
\psfig{figure=1502f2.eps,width=8.8truecm,%
     bbllx=39pt,bblly=37pt,bburx=377pt,bbury=760pt}
\caption[ ]{A small section from the survey, towards $\ell \sim 50\degr$,
 illustrates the calibration of the total intensity data to the
absolute temperature scale. The panels, from top to bottom, display the
Effelsberg (contours start at 100~mK and are plotted in 150~mK steps)
and Stockert 1.4~GHz (contours start at 4500~mK in 100~mK steps, full
beam brightness scale) measurements and the combination of the two,
respectively. Contours for the combined map (bottom) run from 5500~mK
in steps of 150~mK. }
\label{einha}
\end{figure}

\clearpage

We use the modified method only when  there are non-negligible
scanning effects in the source  component of the Stockert data. A
sample region (towards $\ell \sim 50\degr$ ) is given in
Fig.~\ref{einha} to illustrate the effect of the absolute calibration.

\section{Instrumental Polarization}
One major goal of the new 1.4~GHz survey is the sensitive mapping of
polarized emission out of the Galactic plane. However, a problem is
given by the relatively high instrumental polarization mainly
introduced by the cooled broad-band polarization transducer and hybrid
of the L-band Effelsberg receiver. A change of the bandwidth
and centre frequency has a significant effect on the instrumental $U$ and $Q$
components. Other instrumental effects due to the antenna and feed
characteristics, residual ellipticity of the polarimeter response and variations
with time add, but are too small to be separated from the main effect.

An attempt was made to minimize these instrumental effects to a
residual effect of the order of 1\%. We assume that the instrumental
components $U_\mathrm{inst}$ and $Q_\mathrm{inst}$ scale with the total
intensity $I$. When observing in an astronomical coordinate system the
instrumental components depend on the parallactic angle $\phi$. A
procedure was developed to correct for this effect. From the
observations of the polarized calibration sources 3C286 and 3C138
at different parallactic angles, correction factors $f_\mathrm{U}$ and
$f_\mathrm{Q}$ are calculated in such way that the nominal percentage
polarization and the polarization angle are obtained.

The algorithm of the elimination of the instrumental effects is as follows:
Suppose $U_\mathrm{act}$ and $Q_\mathrm{act}$ are the intrinsic values
of a source in an astronomical coordinate system. $U_\mathrm{obs}$ and
$Q_\mathrm{obs}$ are the observed values depending on the parallactic
angle $\phi$. The instrumental values $U_\mathrm{inst}$ and $Q_\mathrm{inst}$
are:
\begin{eqnarray}
\label{eqnuq}
 {U}_\mathrm{inst}(\phi) &=& {U}_\mathrm{obs}(\phi) - {U}_\mathrm{act}\\
\nonumber
 {Q}_\mathrm{inst}(\phi) &=& {Q}_\mathrm{obs}(\phi) - {Q}_\mathrm{act}
\end{eqnarray}

The instrumental polarization angle $PA_\mathrm{inst}$ and polarization
intensity $PI_\mathrm{inst}$ are:
\begin{eqnarray}
 {PA}_\mathrm{inst}(\phi) & = &\frac{1}{2} \tan^{-1} \left[
\frac{{U}_\mathrm{inst}(\phi)} { {Q}_\mathrm{inst} (\phi) } \right] \\
\nonumber
{PI}_\mathrm{inst}(\phi) & = &\sqrt{ {U}_\mathrm{inst}^2 (\phi) +
{Q}_\mathrm{inst}^2 (\phi) }
\end{eqnarray}
The correction factors $f_\mathrm{U}$ and $f_\mathrm{Q}$ are then:
\begin{eqnarray}
\label{fufq}
f_\mathrm{U}(\phi) = \frac{\displaystyle {PI}_\mathrm{inst}(\phi)}
 {\displaystyle{I}_\mathrm{obs}} \sin \left[ 2 (PA_\mathrm{inst}(\phi) -
\phi ) \right] \\
 \nonumber
f_\mathrm{Q}(\phi) = \frac{\displaystyle {PI}_\mathrm{inst}(\phi)}
 {\displaystyle{I}_\mathrm{obs}} \cos \left[ 2 ({PA}_\mathrm{inst}(\phi)
 - \phi ) \right]
\end{eqnarray}

In Fig.~\ref{fufq} we show, as an example, the data for $f_\mathrm{U}$
and $f_\mathrm{Q}$ from observations of 3C286. $f_\mathrm{U}$ and
$f_\mathrm{Q}$ are determined for each session. It is obvious that
$f_\mathrm{U}$ and $f_\mathrm{Q}$ vary with parallactic angle,
which needs to be taken into account when correcting a survey map.
According to the introduced procedure each pixel $I$ of a map is
multiplied with the appropriate $f_\mathrm{U}$ and $f_\mathrm{Q}$.
The resulting $U$ and $Q$ components of the instrumental polarization are
transformed into parallactic angle corrected components for each pixel
of an $U$ and $Q$ map and subtracted from the observed $U$ and $Q$. The
procedure corrects for most of the instrumental effects. The corrected
$U$ and $Q$ maps are then used to calculate the polarized intensity
$PI$ and the polarization angle $PA$.

\begin{figure}[htb]
\psfig{file=1502f3.eps,width=8.8cm,bbllx=66pt,bblly=123pt,bburx=316pt,bbury=327pt}
\caption{Variation of the calculated $f_\mathrm{U}$ and $f_\mathrm{Q}$
factors with respect to the paralactic angle for 3C286 during one
observing session. Error bars are the standard deviations of the
factors.}
\label{facuq2}
\end{figure}

It has been found that the correction effect on the large-scale
polarization emission is not significant. Strong sources, however,
cause distortions which are clearly minimized by the described
procedure. In Fig.~\ref{sou12} we illustrate the effect of our
correction procedure. We note that instrumental effects do not always
act in a way to increase the observed polarization. Depending on the
arrangement and variations in $U$ and $Q$ it is equally possible to
observe the reverse of this effect.

\begin{figure*}[htb]
\hfill{}
\psfig{file=1502f4a.ps,width=18cm,bbllx=50pt,bblly=110pt,bburx=580pt,bbury=395pt,clip=}
\hfill{}

\hfill{}
\psfig{file=1502f4b.ps,width=18cm,bbllx=50pt,bblly=110pt,bburx=580pt,bbury=395pt,clip=}
\hfill{}
\caption{A sample region to illustrate the effect of the procedure to
eliminate the instrumental polarization. The upper panel displays the
same region before (left) and after (right) the correction. Contours
show the total intensity starting from $-$50~mK with 50~mK intervals.
The strong source around $\ell \sim 47\degr$ is 3C386 with a flux
density of 6.5~Jy. Its percentage polarization is 2.7\% after the
procedure is applied. In the lower panel the polarized intensity data
of the same regions are displayed in grey scale. Contour levels run
starting from 10~mK with 30~mK steps. Again the original data are at
left and corrected data are at right. The plotted electric field
vectors are scaled to the polarized intensity and astronomical
coordinates such that 100~mK correspond to a vector of length 1\farcm 5. }
\label{sou12}
\end{figure*}

\section{Procedure for the absolute polarization adjustment}
As was mentioned earlier the large-scale information in the
polarization maps is lost due to the baseline fitting procedure. In
addition, residual ground radiation effects may show up in view of the
large sizes of the observed maps. As in the case of total power maps,
the polarization emission needs to be adjusted to an absolute level.
However, there exists no complete and regularly gridded low resolution
polarization survey which is absolutely calibrated. Fortunately, a
large set of linear polarization observations at 1.4~GHz obtained with
the Dwingeloo 25-m telescope exists (Brouw \& Spoelstra\ \cite{brouw76}).
These data are absolutely calibrated and corrected for all kinds of
instrumental effects and therefore match the requirements for this task.
However, the data are not on a regular grid and are significantly
undersampled. This data set has to be regridded to be used to adjust
our polarization measurements. The polarized intensity and polarization
angle data from the Dwingeloo survey have been provided by
Dr.~Spoelstra in digitized form.

Using the $PI$ and $PA$ values from this measurements we calculated $U$
and $Q$ values and regridded the undersampled Dwingeloo data on the
grid of the Effelsberg maps. However, in some regions data points are
too separated ($2\degr$ or more) for an absolute calibration, and in other weakly
polarized regions an S/N-ratio of 2 or less does not allow a proper adjustment.

We tried two methods to interpolate the Dwingeloo data on the same
grid as the Effelsberg maps. One way to do this is an interpolation
between the data points using a ``cubic-spline interpolation''.
However, we found that a cubic-spline interpolation introduces
distortions at the corners of the maps and data from a much larger area
must be used to avoid this problem. Moreover, a single high intensity
data point affects surrounding low intensity data up to a large
distance. Data points must be weighted with respect to the distance.
Hence, we found that a cubic-spline interpolation is inadequate for
most of our regions.

A successful method is to weigh data points by their distances to the
required map element. We used the approach $\exp(-\alpha~{\cal R})$ in
which $\alpha$ is a constant. We calculate for each pixel of an
Effelsberg map the Dwingeloo $U$ and $Q$ data within a radius, ${\cal
R}$, contributing with a weight as given above. For our case we
found a value of $6\degr$ for ${\cal R}$ with $\alpha=1$ to be satisfactory.

The reconstructed $U$, $Q$ and $PI$ Dwingeloo maps of a test region at the
$4\arcmin$ grid of the Effelsberg maps are given in Fig.~\ref{dwin}.

\begin{figure*}
\hfill{}
\psfig{file=1502f5a.ps,width=11.5cm,bbllx=40pt,bblly=215pt,bburx=600pt,bbury=572pt,clip=}
\hfill{}

\hfill{}
\psfig{file=1502f5b.ps,width=11.5cm,bbllx=40pt,bblly=215pt,bburx=600pt,bbury=572pt,clip=}
\hfill{}

\hfill{}
\psfig{file=1502f5c.ps,width=11.5cm,bbllx=40pt,bblly=215pt,bburx=600pt,bbury=572pt,clip=}
\hfill{}
\caption{Polarization maps reconstructed from the Dwingeloo data as
explained in the text. The panels, from top to bottom, display the
Stokes $U$ and $Q$ maps and polarization intensity, respectively. The
area marked with dashed lines is the region observed with the
Effelsberg telescope and used to demonstrate the method of absolute
calibration for the polarization data. }
\label{dwin}
\end{figure*}

Figure~\ref{efuq} shows the $U$ and $Q$ maps of the original Effelsberg
measurements of an area, which is a small section of the Dwingeloo map
shown in Fig.~\ref{dwin}. In the higher resolution Effelsberg maps
numerous small-scale polarization structures are visible, which are
smoothed out by the large Dwingeloo beam.

\begin{figure*}
\hfill{}
\psfig{file=1502f6a.ps,width=14.0cm,bbllx=45pt,bblly=210pt,bburx=587pt,bbury=580pt,clip=}
\hfill{}

\hfill{}
\psfig{file=1502f6b.ps,width=14.0cm,bbllx=45pt,bblly=210pt,bburx=587pt,bbury=580pt,clip=}
\hfill{}
\caption{Polarization $U$ and $Q$ maps from the Effelsberg 1.4~GHz
survey. Contours are plotted starting from the lowest value of the
wedge and in steps of 30~mK T$_\mathrm{B}$. Contours starting from zero
are plotted in white with the same steps.}
\label{efuq}
\end{figure*}

The two data sets are combined as follows: The Effelsberg map is
convolved to the Dwingeloo beam (36$\arcmin$) and subtracted from the
Dwingeloo map. The difference is added to the original Effelsberg map.
Figure~\ref{efduq} and Fig.~\ref{efdpi} show the combination for $U$
and $Q$ and the corresponding $PI$ map. In these figures the
small-scale structures are much less pronounced due to the addition of
the strong large-scale polarized emission which varies in the
range from 200~mK to 800~mK across the map (Fig. \ref{dwin}).

The absolutely calibrated $U$ and $Q$ maps may be decomposed into
small-scale and large-scale features by standard methods for total
intensity maps and relative $PI$ maps can be calculated. Some examples
have been given by Uyan{\i}ker (\cite{uyaniker97}).

Plots of the polarization vectors demonstrate the effect of adjusting
the Effelsberg data to an absolute level. The polarization angle maps
are presented as vector plots in Fig.~\ref{efdpivec}. Numerous
small-scale structures are visible in the original Effelsberg map. The
polarization angle varies largely across the maps. However, the
electric field vectors are almost constant in the combined
Effelsberg-Dwingeloo map. The data of this map are the same as those
for Fig.~\ref{efdpi}, but grey-scale representation is much more
sensitive to small variations.

\section{Final remarks}
We described the observation and reduction technique for a sensitive
1.4~GHz continuum and polarization survey with the Effelsberg 100-m
telescope. We illustrated by an example the absolute calibration of the
total intensity Effelsberg data by making use of the 1.4~GHz Stockert
survey and showed the effect on the Effelsberg polarization maps after
adjusting the large-scale polarized emission with the Dwingeloo 1.4~GHz
measurements.

The methods used and introduced in this work have been found to be
appropriate to conduct a 1.4~GHz survey which is absolutely calibrated
both in total and polarized intensity and corrected for instrumental
polarization down to a level of 1\%.

\begin{acknowledgements}
We are very grateful to Dr. Titus Spoelstra for providing the
Dwingeloo polarization data in digital form. We thank Dr. W.N. Brouw for
helpful comments on the manuscript.
\end{acknowledgements}

\begin{figure*}
\hfill{}
\psfig{file=1502f7a.ps,width=14.0cm,bbllx=45pt,bblly=210pt,bburx=587pt,bbury=580pt,clip=}
\hfill{}

\hfill{}
\psfig{file=1502f7b.ps,width=14.0cm,bbllx=45pt,bblly=210pt,bburx=587pt,bbury=580pt,clip=}
\hfill{}
\caption{Polarization $U$ and $Q$ maps after calibrating the
Effelsberg polarization maps to absolute temperature scale. Contours
are plotted starting from the lowest value of the wedge and in
steps of 30~mK T$_\mathrm{B}$. For $U$ map the contours starting at
200~mK T$_\mathrm{B}$ and for the $Q$ map the contours starting at
400~mK T$_\mathrm{B}$ are plotted in white. Contour steps are always
30~mK T$_\mathrm{B}$. }
\label{efduq}
\end{figure*}

\begin{figure*}
\hfill{}
\psfig {file=1502f8a.ps,width=14.0cm,bbllx=45pt,bblly=210pt,bburx=587pt,bbury=580pt,clip=}
\hfill{}

\hfill{}
\psfig {file=1502f8b.ps,width=14.0cm,bbllx=45pt,bblly=210pt,bburx=587pt,bbury=580pt,clip=}
\hfill{}
\caption{Polarization intensity maps before (top) and after (bottom)
calibrating the Effelsberg polarization maps to absolute temperature
scale. Contours are plotted starting from the lowest value of the wedge
and in steps of 30~mK T$_\mathrm{B}$. For the upper panel contours
starting at 120~mK T$_\mathrm{B}$ and for the lower
panel contours starting at 450~mK T$_\mathrm{B}$ are plotted in white. }
\label{efdpi}
\end{figure*}

\begin{figure*}
\hfill{}
\psfig{file=1502f9a.ps,width=14.0cm,bbllx=45pt,bblly=210pt,bburx=580pt,bbury=580pt,clip=}
\hfill{}

\hfill{}
\psfig{file=1502f9b.ps,width=14.0cm,bbllx=45pt,bblly=210pt,bburx=580pt,bbury=580pt,clip=}
\hfill{}
\caption{ Electric field vectors before (top) and after (bottom) calibrating the
Effelsberg polarization maps to absolute temperature scale. The lower panel is
essentially the same as Fig.~8.
Due to the high polarization intensity of background polarization small-scale
variations from the Effelsberg observations are hidden. Electric field vectors are
scaled to the polarized intensity such that 100~mK correspond to a vector of
length 3$\arcmin$ for the upper panel and 1$\arcmin$ for the lower panel. }
\label{efdpivec}
\end{figure*}

\end{document}